\documentclass{PoS}

\title{NNLO W Pair Production at the LHC: Status Report}

\ShortTitle{NNLO W Pair Production at the LHC: Status Report}

\author{\speaker{Grigorios CHACHAMIS}%
         \thanks{The speaker wishes to thank the
         Institut f\"ur Theoretische Teilchenphysik und Kosmologie, RWTH Aachen,
         for hospitality during the latest stages of work on this
         project.}\\
        Instituto de F\'{\i}sica Corpuscular, Universitat de Val\`encia -- 
Consejo Superior de Investigaciones Cient\'{\i}ficas,
Parc Cient\'{\i}fic, E-46980 Paterna (Valencia), Spain\\
        E-mail: \email{grigorios.chachamis@ific.uv.es}}

\abstract{We discuss the result with full mass dependence for the virtual 
NNLO QCD corrections to the W boson pair production in the quark-anti-quark 
annihilation channel. We also report on our progress regarding the treatment 
of the double-real radiative corrections which, along with the virtual-real corrections, 
are the other two necessary ingredients for a theoretical prediction of the total cross 
section for W$^+$ W$^-$ production to NNLO accuracy.}

\FullConference{The European Physical Society Conference on High Energy Physics \\
                 18-24 July, 2013\\
                 Stockholm, Sweden}

\begin{document}

\section{Introduction}

Diboson production plays a crucial
role in the Large Hadron Collider (LHC) physics program.
Any deviations of the measured cross sections
from the Standard Model (SM) predictions would indicate new physics.
The deviations could arise from anomalous triple vector boson interactions or from 
new particles decaying into gauge bosons. Furthermore,
the measurements of WW 
and ZZ production cross sections are essential for Higgs boson studies.

In particular, W pair production in the quark-anti-quark annihilation channel,
\begin{equation}
q {\bar q} \rightarrow \mathrm{W}^+ \, \mathrm{W}^- \, ,
\label{chachamis_qq_channel}
\end{equation}
plays an essential role as it serves as a signal process 
in the search for new physics and also is the dominant irreducible
background to the Higgs poduction channel 
$p p \rightarrow \mathrm{H} \rightarrow \mathrm{W}^* \mathrm{W}^* \rightarrow l {\bar \nu} {\bar l}' \nu' $.
Both ATLAS~\cite{ATLAS:2012mec} and CMS~\cite{ Chatrchyan:2013oev} collaborations have released measurements of the $\mathrm{WW}$ cross section
at $\sqrt{s} = 7$ TeV whereas CMS has also
published data for $\sqrt{s} = 8$ TeV~\cite{Chatrchyan:2013yaa}.
The measured cross section for W boson pair production, by both collaborations,
shows a tendency to be larger than
the current theoretical prediction within the SM, a fact that calls in itself 
for more accurate theoretical predictions.

The process~(\ref{chachamis_qq_channel}), initially calculated at leading order
more that three decades ago~\cite{Brown:1978mq}, 
 is currently known at next-to-leading order (NLO) 
accuracy in 
QCD~\cite{Ohnemus:1990za, Baur:1995uv, Campbell:1999ah, Frixione:2002ik, 
Hamilton:2010mb, Billoni:2013aba},
after a tremendous effort.
Soft-gluon resummation effects have been assessed 
in Refs.~\cite{Grazzini:2005vw, Dawson:2013lya} while
the electroweak corrections are also known to NLO~\cite{Accomando:2004de, Kuhn:2011mh, Bierweiler:2012kw, Baglio:2013toa}.
The NLO QCD corrections were proven to be large, 
enhancing the tree-level result by
almost 70\% which falls to a still large 30\% after 
imposing a jet veto. 
Therefore, for a more accurate theoretical estimate --with un uncertainty of less than 10\%--
to be compared against the
experimental measurements at the LHC, one is bound to 
go one order higher in the perturbative expansion, namely,
to the next-to-next-to-leading order (NNLO).

To complete the picture, there is another process that needs to be considered and which
is the W pair production
in the loop-induced gluon fusion channel, 
\begin{equation}
g g \rightarrow \mathrm{W}^+ \mathrm{W}^- \, .
\label{chachamis_gg_channel}
\end{equation}
The latter
contributes at $\mathcal{O}(\alpha_s^2)$ relative to the 
quark-anti-quark-annihilation channel but is 
nevertheless enhanced due to the large gluon flux
at the LHC~\cite{Dicus:1987dj, Binoth:2005ua}.

Beyond NLO accuracy in QCD, there are only partial results in the 
literature~\cite{Chachamis:2007cy, Chachamis:2008xu, Gehrmann:2013cxs, Campanario:2013wta}.
We have computed the NNLO two-loop~\cite{Chachamis:2007cy}
and the one-loop squared~\cite{Chachamis:2008xu}
virtual corrections in the high energy limit few years ago.
However, this is not enough to cover
the kinematical region close to threshold. Therefore,
in order to cover all kinematical regions we proceed as follows.
We perform a deep expansion in the W mass around the
high energy limit which in combination with
the method
of numerical integration of differential
equations~\cite{Caffo:1998du, Boughezal:2007ny, Czakon:2007qi} 
allows us the numerical computation of the two-loop\footnote{We work 
in a similar fashion for the 
one-loop squared virtual corrections. However, here we restrict the discussion to the two-loop
corrections only.}
amplitude
with full mass dependence over the whole phase space.
We report our progress on that as well as on the computation
of the double-real NNLO corrections in the following Sections.

\section{The virtual NNLO corrections}

The method for computing the amplitude in the
high energy limit, namely the limit where $s$ is
much larger than the W mass, $m$, 
is similar to the one followed in Refs.~\cite{Czakon:2007ej, Czakon:2007wk}.
The amplitude is reduced such 
that it only contains a small number of integrals (master integrals) 
by use of the Laporta algorithm~\cite{Laporta:2001dd}. 
Next comes the derivation, in a fully 
automatized way, of the Mellin-Barnes (MB) 
representations~\cite{Smirnov:1999gc, Tausk:1999vh} 
of all the master integrals by using the MATHEMATICA package
\textsc{MBrepresentation}~\cite{MBrepresentation}. 
The MB representations are in turn analytically continued in the number 
of space-time dimensions by means
of the {\bf MB} 
package~\cite{Czakon:2005rk}, thus revealing the full singularity 
structure. An asymptotic expansion in
the mass is performed by closing contours and the 
integrals are finally resummed, either
with the help of \textsc{XSummer}~\cite{Moch:2005uc} or the 
\textsc{PSLQ} algorithm~\cite{pslqAlg}.
The final result is expressed in terms of harmonic polylogarithms.

The high energy limit result by itself is not
enough, as was mentioned before. 
The next step, following the methods applied in Ref.~\cite{Czakon:2008zk},
is to compute power corrections in the W mass. 
Power corrections are good enough to
cover most of the
phase space, apart from the region near 
threshold as well as the regions corresponding
to small angle scattering.

We introduce here some of the notation of Ref.~\cite{Chachamis:2007cy}
for completeness.
The W pair production process in the quark-anti-quark annihilation channel reads in detail:
\begin{equation}
\label{chachamis_qqWW}
q(p_1) + {\overline q}(p_2) 
\:\:\rightarrow\:\: \mathrm{W}^-(p_3,m) + \mathrm{W}^+(p_4,m) \, ,
\end{equation}
where $p_i$ denote 
the quark and W momenta.

We choose to express the amplitude in terms
of the kinematic variables $x$ and $m_s$ which are defined to be
\begin{equation}
  x = -\frac{t}{s}, \;\; m_s = \frac{m^2}{s},
\end{equation}
where
\begin{equation}
s = (p_1+p_2)^2 \;\; {\rm and} \;\;t = (p_1-p_3)^2-m^2\,.
\end{equation}
The variation then of
$x$ within the range $ [ 1/2(1-\beta), 1/2(1+\beta) ] $, where
$\beta=\sqrt{1-4m^2/s}$ is the velocity, corresponds to angular variation
between the forward and backward scattering.

Any master integral $M_i$ can be written then as
\begin{equation}
M_i = M_i \left( m_s, x, \epsilon \right) = \sum_{j=k}^l \epsilon^j {I_i}_j(m_s, x),
\label{chachamis_epReveal}
\end{equation}
where $\epsilon$ is the usual regulator in dimensional regularization ($d = 4 - 2 \epsilon$)
and the lowest power of $\epsilon$ in the sum can be $-4$.

The key point now, is that
the derivative of any Feynman integral with
respect to any kinematical variable is 
again a Feynman integral with denominators or numerators 
raised to some new powers and which may be
reduced anew in terms of the initial set of master integrals.
This way, one can construct a
partially triangular system of differential equations in $m_s$,
which can subsequently be solved in the form of a power series expansion
in $m_s$.

If we differentiate an arbitrary master integral, $M_i(m_s,x,\epsilon)$, 
with respect to $m_s$ and $x$, we will have
respectively
\begin{equation}
m_s \frac{d}{dm_s} M_i(m_s,x,\epsilon) = 
\sum_j C_{i j}(m_s,x,\epsilon) ~M_j(m_s,x,\epsilon)
\label{chachamis_dms}
\end{equation}
and
\begin{equation}
x \frac{d}{dx} M_i(m_s,x,\epsilon) = 
\sum_j C_{i j}\sp{\prime}(m_s,x,\epsilon) ~M_j(m_s,x,\epsilon)\, .
\label{chachamis_dx}
\end{equation}
We use Eq.~(\ref{chachamis_dms}) to obtain the 
mass corrections for the master integrals  
calculating the power series expansion
up to order $m_s^{30}$ (see also Ref.~\cite{Czakon:2008zk} for more
details).
This deep expansion in $m_s$ should be sufficient for most
of the phase space but still not enough to
cover the whole allowed kinematical region. From this point on, the way to
proceed is by numerically integrating
the system of differential equations.
We choose to work with
the master integrals in the form of Eq.~(\ref{chachamis_epReveal}), where the
$\epsilon$-dependence is explicit. We can then work with the coefficients
of the $\epsilon$ terms and accordingly have
\begin{equation}
m_s \frac{d}{dm_s} I_i(m_s,x) = 
\sum_j J^M_{i j}(m_s,x) ~I_j(m_s,x)
\label{chachamis_I_dms}
\end{equation}
and
\begin{equation}
x \frac{d}{dx} I_i(m_s,x) = 
\sum_j J^X_{i j}(m_s,x) ~{I_j\sp{\prime}}(m_s,x),
\label{chachamis_I_dx}
\end{equation}
where the Jacobian matrices $J^M$ and $J^X$ have as elements rational
functions.

By using the system of differential equations~(\ref{chachamis_I_dms},~\ref{chachamis_I_dx}),
one can obtain a 
numerical solution for the amplitude. 
Actually, what we are dealing now with is an initial
value problem and the main requirement is 
to have the initial conditions to high enough accuracy. 
The initial conditions are nothing but the values of the master integrals
at a proper kinematical point, which we call initial
point, and can be provided by the power series expansion.
The initial point has to be chosen somewhere
in the high energy limit region, where
$m_s$ is small and therefore, the values obtained by the power series
are very accurate.
Starting from there, one can evolve to any other point
of the phase space after integrating numerically the system of differential
equations~(\ref{chachamis_I_dms},~\ref{chachamis_I_dx}).

We parametrize with a suitable grid of points the region close
to threshold and then we calculate the master integrals for the points of the
grid by evolving as described above.
Given that the master integrals have to be
very smooth, one can use interpolation,
after having the values for the grid points,
in order to get the values at any point of the region.
We use 1600 points for the grid and take as initial conditions
the values of the master integrals at the
point $m_s = 5 \times 10^{-3}$, $x = 1/4$. 

The numerical integration is performed by using \textsc{Odepack}~\cite{ODEPACK},
one of the most advanced
software packages implementing the variable coefficient multistep method. 
We use quadruple precision to maximize
accuracy.
The values at any single grid  point can be obtained
in about 10 minutes in average (with a typical 2.5GHz Intel Core 2 Duo system)
after compilation with the \textsc{Intel Fortran Compiler}.
The achieved accuracy is beyond 10 digits for most of the points
of the grid.
It is also worth noting that in order to perform the numerical
integration one needs to deform the contour in the complex plane
away from the real axis. This is due to the fact
that along the real axis there are spurious
singularities. 
We use an elliptic contour and 
we achieve a  better estimate of the final global error
by calculating more than once for each point of the grid, 
using each time different eccentricities.
We will not present here any numbers since the aim was to report on the
general methods. The details and the results of the study will be  presented in a future 
publication~\cite{chachamis_czakon}. 

\section{The double-real NNLO corrections}

In this section we will only describe our methodology as this is ongoing work, the results
will be given in a following publication.

To begin with, the matrix elements for the double-real NNLO QCD corrections 
are needed and these can be obtained
by the following processes:
\begin{eqnarray}
\label{chachamis_qqWWgg}
0 \rightarrow u(p_1) + {\overline u}(p_2)  +  \mathrm{W}^-(p_3,m) + \mathrm{W}^+(p_4,m) +
 g(p_5) + g(p_6)\, ,\\
 0 \rightarrow u(p_1) + {\overline u}(p_2)  +  \mathrm{W}^-(p_3,m) + \mathrm{W}^+(p_4,m) +
q\sp{\prime}(p_5) + {\overline q}\sp{\prime}(p_6)  \, , \\
0 \rightarrow d(p_1) + {\overline d}(p_2)  +  \mathrm{W}^-(p_3,m) + \mathrm{W}^+(p_4,m) +
 g(p_5) + g(p_6)\, ,\\
 0 \rightarrow d(p_1) + {\overline d}(p_2)  +  \mathrm{W}^-(p_3,m) + \mathrm{W}^+(p_4,m) +
q\sp{\prime}(p_5) + {\overline q}\sp{\prime}(p_6)  \, ,
\end{eqnarray}
where $u$ denotes an up-type quark and $d$ a down-type one, 
if we assume all possible crossings of quarks and
gluons in the initial state, such that we end up with a $2 \rightarrow 4$
process while the W pair remains in the final state.
Similarly, the NLO real
matrix elements are given by crossings to the processes:
\begin{eqnarray}
\label{chachamis_qqWWg}
0 \rightarrow u(p_1) + {\overline u}(p_2)  +  \mathrm{W}^-(p_3,m) + \mathrm{W}^+(p_4,m) +
 g(p_5) \, ,\\
0 \rightarrow d(p_1) + {\overline d}(p_2)  +  \mathrm{W}^-(p_3,m) + \mathrm{W}^+(p_4,m) +
 g(p_5) \, ,
 \end{eqnarray}
whereas the LO one is given by Eq.~(\ref{chachamis_qqWW}) for q being either an up-type
quark or a down-type one.
We generate the matrix elements using \textsc{FeynArts}~\cite{Hahn:2000kx}, we simplify
them with \textsc{Form}~\cite{Vermaseren:2000nd} and, as a cross-check, we compare 
against \textsc{MadGraph}~\cite{Alwall:2011uj}.

With the matrix elements at hand, we use
the general subtraction scheme \textsc{STRIPPER}~\cite{Czakon:2010td}
for the evaluation of the NNLO QCD contributions 
from double-real radiation. The result is a Laurent expansion in the parameter of dimensional regularization, the coefficients of which are evaluated by numerical Monte Carlo integration. 
The two main ideas behind  \textsc{STRIPPER}
 are a two-level decomposition of the phase space, the second one factorizing the singular limits of amplitudes, and a suitable parameterization of the kinematics allowing for derivation of subtraction and integrated subtraction terms from eikonal factors and splitting functions without non-trivial analytic integration.

\section{Conclusions}
The production of a W$^+$ W$^-$ pair via quark-anti-quark-annihilation
is an important process in the LHC physics program for which higher
accuracy theoretical estimates are essential. 
After having calculated
the two-loop and the one-loop-squared virtual QCD corrections
to the W boson pair production in the high energy limit, we described how we use
a combination of a deep expansion in the W mass around the
high energy limit and numerical integration of differential
equations to compute the virtual amplitudes
with full mass dependence over the whole phase space. 
Finally, we presented our methodology for evaluating 
the double-real NNLO QCD corrections.
\section*{Acknowledgements}

This work has been supported by Marie Curie Actions (PIEF-GA-2011-298582)
and in part by the Spanish Government and ERDF funds from the EU Commission 
[Grants No. FPA2011-23778, No. CSD2007-00042 (Consolider Project CPAN)] 
and by Generalitat Valenciana under Grant No. PROMETEOII/2013/007.

\end{document}